\documentclass[twocolumn,aps,prl,
superscriptaddress
]{revtex4-2}

\usepackage{graphicx,color,hyperref}

\usepackage{amsmath}
\usepackage{amsfonts}
\usepackage{amssymb}

\usepackage{braket}

\begin{document}

\title{Experimental Observation of Topological Quantum Criticality}

\author{Sonja Barkhofen}
\affiliation{Integrated Quantum Optics Group, Institute for Photonic Quantum Systems (PhoQS), Paderborn University, Warburger Stra\ss{}e 100, 33098 Paderborn, Germany}

\author{Syamsundar De }
\affiliation{Integrated Quantum Optics Group, Institute for Photonic Quantum Systems (PhoQS), Paderborn University, Warburger Stra\ss{}e 100, 33098 Paderborn, Germany}
\affiliation{Advanced Technology Development Centre, Indian Institue of Technology Kharagpur, Kharagpur 721302, India}

\author{Jan Sperling}
\affiliation{Integrated Quantum Optics Group, Institute for Photonic Quantum Systems (PhoQS), Paderborn University, Warburger Stra\ss{}e 100, 33098 Paderborn, Germany}

\author{Christine Silberhorn}
\affiliation{Integrated Quantum Optics Group, Institute for Photonic Quantum Systems (PhoQS), Paderborn University, Warburger Stra\ss{}e 100, 33098 Paderborn, Germany}

\author{Alexander Altland}
\affiliation{Institut f\"ur Theoretische Physik, Universit\"at zu K\"oln, Z\"ulpicher Stra\ss e 77, 50937 K\"oln, Germany}

\author{Dmitry Bagrets}
\affiliation{Institut f\"ur Theoretische Physik, Universit\"at zu K\"oln, Z\"ulpicher Stra\ss e 77, 50937 K\"oln, Germany}

\author{Kun Woo Kim}
\affiliation{Department of Physics, Chung-Ang University, 06974 Seoul, Republic of Korea}

\author{Tobias Micklitz}
\affiliation{Centro Brasileiro de Pesquisas F\'isicas, Rua Xavier Sigaud 150, 22290-180, Rio de Janeiro, Brazil }

\date{\today}

\begin{abstract}
We report on the  observation of quantum criticality forming at the transition point between  topological Anderson insulator phases in a one-dimensional photonic quantum walk with spin. The walker's  probability distribution reveals a time-staggered profile of the dynamical spin-susceptibility, recently suggested as a 
smoking gun signature for topological Anderson criticality in the chiral symmetry class AIII. Controlled breaking of phase coherence removes the signal, revealing its origin in quantum coherence.
\end{abstract}

\maketitle

{\it Introduction:---}The presence of disorder in one-dimensional ($1d$) systems generically causes  Anderson localization of  single particle states at microscopically short length scales~\cite{anderson1958absence,Abrahams1979}. The single known exception to this rule is  quantum criticality between different symmetry protected topological phases~\cite{Evers2008,AltlandBagretsKamenev2015}. At criticality, the number of topological boundary states changes, and the only way to do so is by hybridization through the bulk. This topologically enforced delocalization trumps Anderson localization and leads to the transient formation of an extended quantum critical state whose exotic properties include the extremely (logarithmically) slow spreading of wave packages, or vanishing typical (but finite ensemble averaged) conductance~\cite{balents1997delocalization}. In this paper, we report on the experimental observation of such criticality between topological  phases in symmetry class AIII.

The experimental realization of this setting is  challenging. It requires precision control over an internal degree of 
freedom, or `spin', difficult to achieve in ultracold atom setups, otherwise tailored to the observation of 
Anderson localization~\cite{Chabe2008,Hainaut2018,Billy2008}. 
 Second, the identification of reluctant (logarithmically slow) delocalization appears to  require signal observation over exponentially 
long time scales. 

However, the toolbox of quantum optics experimentation  turns out sufficiently versatile to overcome these challenges.   
In a recent work, we proposed the blueprint of a photonic quantum simulator of the 
extended state, and 
a delicate time-staggered signature in the spin susceptibility as smoking gun evidence for criticality already on short time scales \cite{bagrets2021probing}. 
We here report on
the experimental realization of this 
proposal within a tunable optical linear network.

\begin{figure}\centering
  \includegraphics[width=0.9\columnwidth] {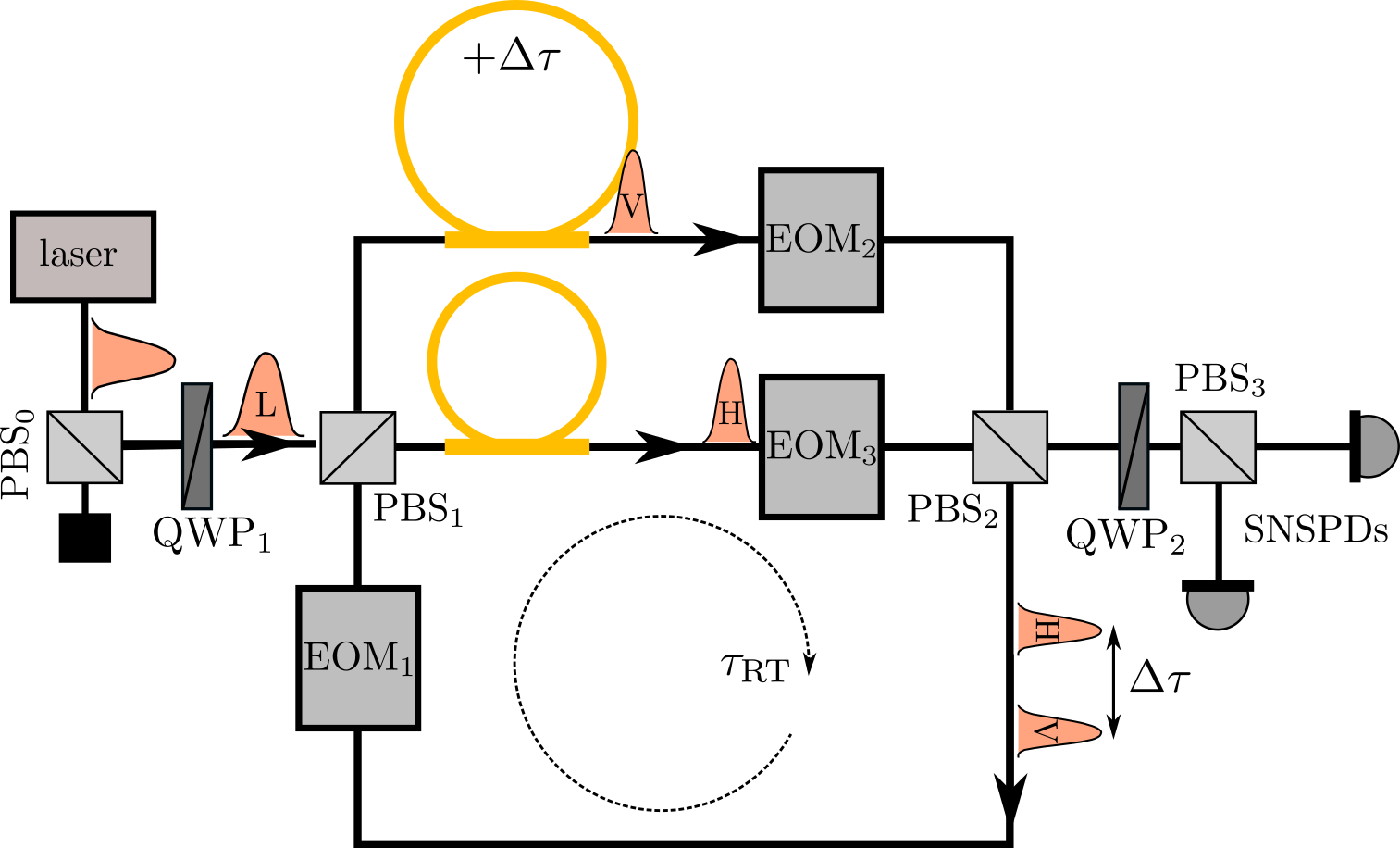} 
  \caption{Experimental setup based on an unbalanced Mach-Zehnder interferometer with dynamic coin and deterministic in/out coupling (see text for details).} \label{Fig:Setup}
\vspace{-.2cm}
\end{figure}

{\it Quantum walk protocol:---}A schematic of the photonic quantum simulator  is shown in Fig.~\ref{Fig:Setup}.
The optical linear network simulates a one-dimensional quantum walk of a spin-$1/2$ particle 
with single time-step Floquet evolution, 
\begin{align}
\label{eq:quantum_walk_protocol}
\hat U(\hat \varphi,\hat\theta)
  &=
    R_3(\tfrac{\hat \varphi}{2}) R_1(\tfrac{\hat \theta}{2})
    \,\hat T \,
    R_1(\tfrac{\hat \theta}{2}) R_3(\tfrac{\hat \varphi}{2}),
\end{align} 
and step- and coin-operators, 
\begin{align}  \label{eq:step}
\hat T 
    &= \sum_q \left( 
    |q+1, \uparrow \rangle  \langle \uparrow, q|
    +
    |q-1, \downarrow \rangle  \langle  \downarrow, q |
    \right),\\  \label{eq:rotation}
R_j(\hat\alpha)
    &=
    \sum_{q, \sigma\sigma' 
    } 
    |q, \sigma \rangle  \left[ 
    e^{-i\alpha_q \hat \sigma_i} \right]_{\sigma \sigma'}\langle q,\sigma'|,\quad j=1,3.
\end{align}
  Here the sums are over lattice sites (with unit spacing) $q \in \mathbb{Z}$, and 
  spin-orientations $\sigma,\sigma'\in \{ \uparrow , \downarrow \}$ parametrizing 
  the walker's internal degrees of freedom, with Pauli matrices $\hat \sigma_i$, $i = 1,2,3$ operating 
  on the latter. Throughout the work we denote eigenstates of $\hat \sigma_3$ 
  and $\hat \sigma_2$ by
$\uparrow , \downarrow$ and $+$, $-$, respectively. 
The latter will play a major role, and are also referred to as circular right ($R$)/left ($L$) polarized states.  Static disorder is introduced by drawing site-dependent angles 
$\varphi_q$, $\theta_q$ from some distribution (see also below), which leaves the average values $\bar{\varphi}$, $\bar{\theta}$
as  tuning parameters.

\begin{figure*}
\centering
\includegraphics[width=2\columnwidth] {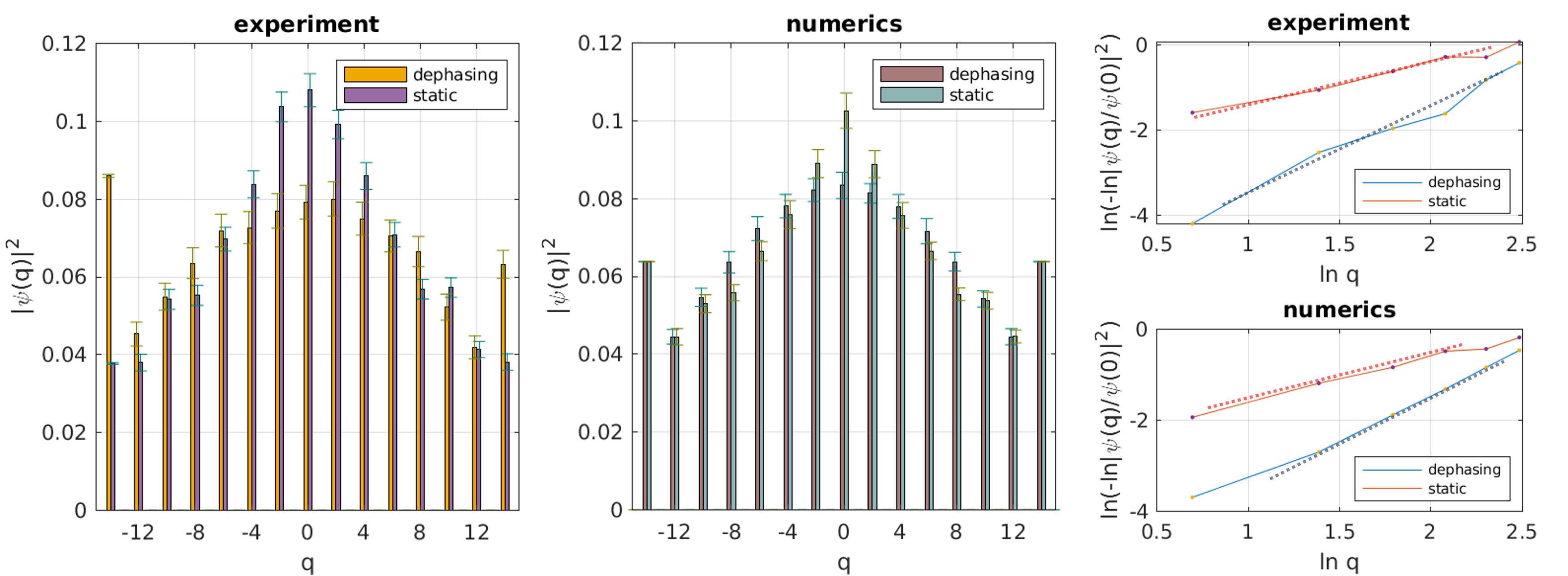}   
  \caption{Experimental (left panel) and numerical (center panel) probability distributions for static (purple/red) and dephasing (orange/brown) disorder.   The distributions are averaged over 500 disorder realizations detected in  the circular $R/L$ basis with  circular $R$ input state,   $\gamma = \theta_\mathrm{EOM}= \pi/8$ for step 14. 
    Right panel: 
    Both, in experiment and numerical simulation, distributions are well described by exponential profiles for static, respectively, Gaussian profiles for dephasing disorder. Dotted lines with slope 1 (red) and 2 (blue) here are introduced for comparison.
  }
  \label{Fig:wavefunction}
  \vspace{-.5cm}
\end{figure*}

The operator $\hat U$ possesses the (chiral) symmetry $\hat\sigma_2 \hat{U}\hat\sigma_2= \hat{U}^\dagger$, putting it into class AIII of the classification scheme.
Eigenstates, $|\psi\rangle$, of Floquet systems with chiral symmetries come in pairs with quasi-energies $\epsilon_0\pm \omega$, mirror symmetric around center energies, $\epsilon_0$. Presently, we have four of those, $\epsilon_0=0,\pi,\pm\pi/2$. The first emerges as  a direct consequence of chirality:  if $|\psi\rangle$ is a state with quasi-energy $\epsilon$, then $\hat\sigma_2|\psi\rangle$ is the one with energy $-\epsilon$. The remaining three originate in a second symmetry, our walk operator anticommutes with $\hat S\equiv 
    \sum_q |q\rangle(-1)^q\langle q|$, which is to say that it hops between neighboring sites, 
     $\ldots  q\longleftrightarrow (q+1)\longleftrightarrow (q+2) \ldots$. It is straightforward to check that $\hat S|\psi\rangle$ is an eigenstate with energy $\epsilon+\pi$\footnote{
We have 
$
    U (\hat\sigma_2 \ket \psi ) = \hat\sigma_2 U^\dagger \ket \psi = e^{-i\epsilon} \hat\sigma_2 \ket \psi$, and $
    U\hat S\ket\psi  = -\hat S U \ket \psi = e^{i\pi + i\epsilon} \ket\psi
$ 
}. Hence, the entire spectrum is $\pi$-shift invariant, explaining the second mirror symmetry $0\to \pi$. The remaining pair is best understood by defining an auxiliary operator $V=i U$. With $(\hat\sigma_2 \hat S) V (\hat\sigma_2 \hat S)=V^\dagger$, it follows that $0$ and $\pi$ are mirror energies of $V$. However, $V$ and $U$ have the same eigenstates, with quasi-energies shifted by a factor $i=e^{i\pi/2}$. This explains why the $(0,\pi)$ pair of $V$ becomes the $(-\pi/2,\pi/2)$ pair of our operator~$U$. 

At critical points separating topologically distinct Anderson insulating phases, 
 gaps in the quasienergy spectrum of the clean system
close and delocalized states at the above quasienergies emerge. 
Considering the walk \eqref{eq:quantum_walk_protocol}, we find that it contains a critical point for the average disorder value $\bar\theta=0$ \cite{bagrets2021probing}  at which all four energies simultaneously are critical \footnote{The associated topological invariants may be identified by analysis of the auxiliary `Hamiltonian' $\hat H=-i\ln \hat U$. However, we will not need this underlying structure~\cite{asboth2012symmetries,asboth_bulk-boundary_2013} throughout.}. Finite constant values $\bar \theta$ gap out the pair $(0,\pi)$ while a staggered configuration $\theta_{\mathrm{s}}= \mathrm{const.}\times (-)^q$ gaps out  $(-\pi/2,\pi/2)$. However, throughout we will keep the system at  $\bar \theta=0$ and identify signatures of the esnuing delocalized states.

{\it Time-staggered spin-polarization:---}Anderson critical states retain `memory' of the anti-unitary symmetries defining them~\cite{balents1997delocalization}: consider the average probability for a walker initially prepared on site $q=0$ in 
spin-state $\sigma$ to be found after $t$ time-steps at a distance $q$ in 
 spin-state $\sigma'$, 
  \begin{align}
  \label{eq:walker_probability}
    P_{\sigma'\sigma} (t,q) 
    &=
    \langle 
    |\langle q,\sigma' |\hat U^t |0,\sigma\rangle |^2
    \rangle_{\theta
    }.
  \end{align}
Our linear optical network 
gives direct experimental access to this observable, where the average $\langle \dots \rangle_{\theta}$ is  over multiple runs, each for a randomly drawn binary configuration $\theta_q\in\{\pm\theta\}$ at constant $\varphi_q=0$. 
Preparing the walker  
in a $\hat\sigma_2$-eigenstate $\sigma\in\{+,-\}$, 
the probabilities 
$P_{+,\sigma}$ and $P_{-,\sigma}$ differ~\cite{bagrets2021probing}, reflecting the origin of criticality in a symmetry involving $\hat\sigma_2$. 
This asymmetry, absent in Anderson localized phases,  motivates the introduction of the spin-polarization
  \begin{align}  \label{eq:integrated_spin_polarization}
    \Delta P(t)\equiv \sum_q 
   \left(
   P_{--}(t,q)-P_{+-}(t,q)
   \right), 
  \end{align} 
 sampled over all sites
as a unique diagnostic of topological quantum criticality.

What makes the measurement of $\Delta P$ challenging is that all states in the quasienergy spectrum contribute to $P$, while only the states of distinguished quasi-energies $(0,\pi)$ and $(-\pi/2,\pi/2)$ contribute to the asymmetry~\footnote{ 
To probe only critical states, one may prepare initial states extended over several sites with fixed phase relations, as discussed in Ref.~\cite{bagrets2021probing}. Their experimental realization is, however, challenging and we thus focus on initial states localized on single sites, probing the entire quasi-energy spectrum.}. More precisely, the states defined by 
$\hat\sigma_2$ yield a signal with smooth time dependence, while those with associated to $\hat\sigma_2\hat S$ yield a staggered signal $\Delta P(t)=(-1)^t |\Delta P(t)|$, corresponding to the sign alternation of the $\hat S$-operator as introduce above \cite{bagrets2021probing}. The added contribution of all states is thus expected to show spectral peaks at $\omega =0,\pi$ in the Fourier transform of $\Delta P(t)$. This is our principal experimental signature of AIII topological quantum criticality in the walk.

\begin{figure*}[t]
\centering \label{fig:main_result}
\includegraphics[width=.9\columnwidth] {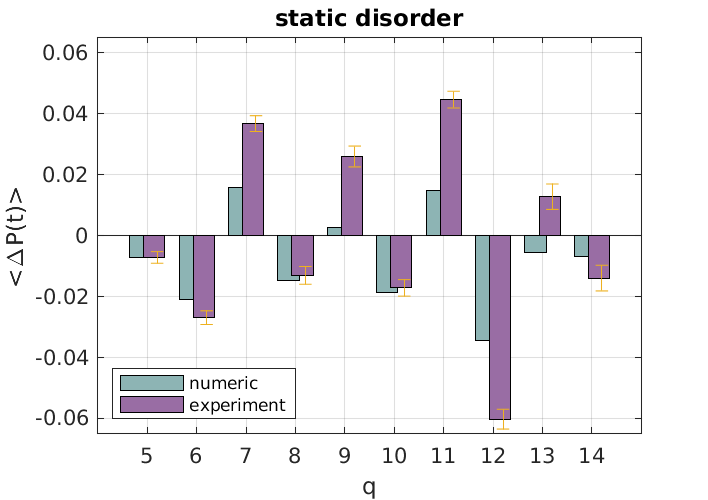}
\includegraphics[width=.9\columnwidth] {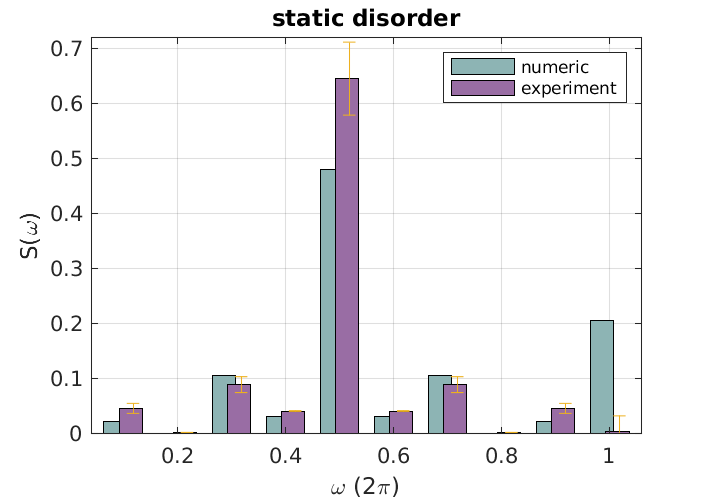}
\includegraphics[width=.9\columnwidth] {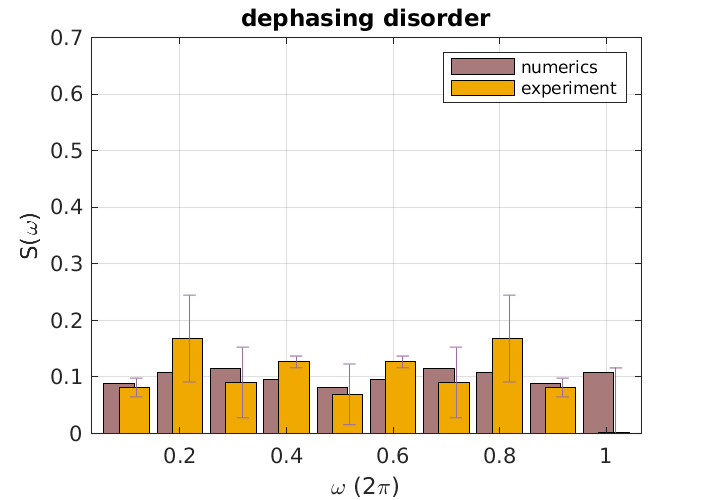}
\includegraphics[width=.9\columnwidth] {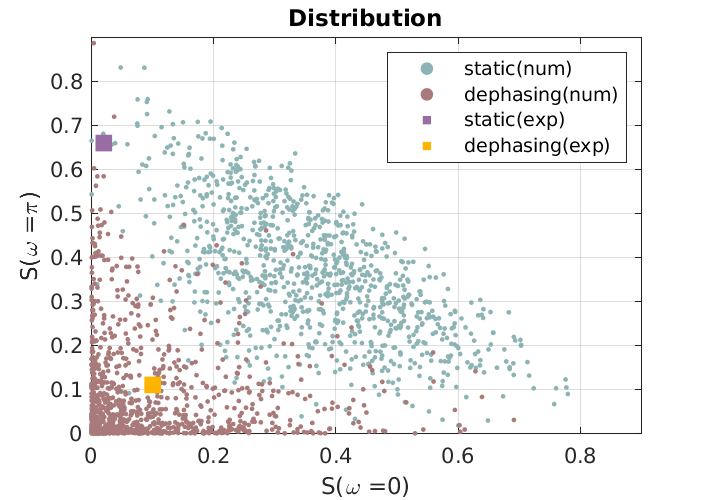}
 \hspace{.2cm}
 \caption{ Comparison of experimental results 
 and numerical simulations
for critical quantum walks with static (upper panels) and dephasing (left-bottom panel) disorder and up to $t=15$ time steps.
Results are averaged over 500 different 
realisations of the quantum walk 
and error bars indicate the statistical error of the 500 measurements. 
 Top-left: 
  Average spin-polarization $\langle \Delta P(t)\rangle$. 
     Top-right: Corresponding power spectra, $S(\omega)$, of  the static binary disorder. 
     Bottom-left: The power spectra of quantum walk with dephasing disorder are shown. 
     Bottom-right: The distribution of power spectrum $S(\omega=0)$ and $S(\omega=\pi)$ is plotted by sampling 1000 points where each data point is obtained from the spin polarization averaged over 500 random disorder realizations. 
     \newline
   \label{Fig:spinPol} \\
\vspace{-.2cm} }
\end{figure*}

{\it Simulator and experimental realization:---}Fig.~\ref{Fig:Setup} shows the experimental setup for our quantum simulator. 
At its core  is a Mach-Zehnder interferometer with a feedback loop realising the chiral translation $\hat T$,   Eq.~\eqref{eq:step},  
via time-multiplexing: A  laser pulse is split into horizontal and vertical polarization orientations 
(the `spin'-components $\uparrow,\downarrow$) 
and send through fiber lines of different lengths. In this way positions on the lattice are mapped onto time domain, with time delay between vertically/horizontally polarized pulses defining the lattice constant~\cite{schreiber_photons_2010,schreiber_decoherence_2011, nitsche2018probing}. The network architecture offers full control over the dynamic coin operations $\hat R$, Eq.~\eqref{eq:rotation} via polarization rotations~\cite{schreiber_decoherence_2011, barkhofen2017measuring}, and allows to measure the spin resolved probability distributions $P_{\sigma'\sigma}$ defined in Eq.~\eqref{eq:walker_probability} \cite{barkhofen2018supersymmetric}. It thus provides the key ingredients of our proposal. 
In the following, we discuss the concrete realization of this protocol in the quantum optical simulator. Readers primarily interested in results are invited to continue reading in the following section. 

We realize the
quantum walker by a weak coherent laser pulse at telecom wavelength and its polarisation acts as the internal degree of freedom.
The initial polarisation is set at quarter wave plate QWP$_1$ at -45$^\circ$, such that left circular light $L$ enters the setup.  
The dynamic coin operation is accomplished by an EOM, implementing the voltage dependent polarisation rotation
\begin{equation}
R_1(\theta_U) = \begin{pmatrix}
  \cos(\theta_U) & -i\sin(\theta_U) \\
  -i\sin(\theta_U)& \cos(\theta_U)
\end{pmatrix}~.
\end{equation}
As the EOM switches fast enough to address each pulse (i.e. each quantum walk position) individually it is capable of  realising both  static and dynamic disorder.
Its programmability makes the recording of hundreds of patterns in short time  without manual setup of parameters possible.
After the coin operation, the light pulses are split  according to their polarisation at a polarising beam splitter (PBS$_1$) and are routed into the two arms of the interferometer. 
The pulses propagating in the upper arm are retarded by a time delay of $\Delta \tau \approx 105$\,ns relative  to those  in the lower arm. 
 The conditional routing of the photons through the long or short fibre realises the chiral translation operator, Eq.~\eqref{eq:step}. The
 time delay $\Delta \tau$ and roundtrip time $\tau_\mathrm{RT}$
here define the 
 lattice spacing and single time step duration, 
 respectively.   
 After PBS$_2$ the pulses are feedbacked in the loop and the dynamics continues until the desired final step, in which the train of pulses is deterministically coupled out by the EOMs 2 and 3.
Given that these EOMs flip the polarization for the in and outcoupling the full dynamics of an $N$ step evolution is described by 
\begin{eqnarray*}
  \hat U_\mathrm{full} = \hat\sigma_1 (T^{-1}   R_1(\theta_U))^N 
  \hat\sigma_1 T   .
\end{eqnarray*}
The application of the operator $T$ describes the first transition through the fibre arms in which the role of the horizontally and the vertically polarized light is swapped with respect to the following roundtrips.
The consecutive $\hat\sigma_1$ matrices signify the polarization flip by the EOMs 2 and 3 for in and outcoupling.
As $\hat\sigma_1$ commutes with $R_1$ and using $T = \hat\sigma_1 T^{-1}\hat\sigma_1$  we obtain for the full evolution
\begin{eqnarray*}
  \hat U_\mathrm{full} =  (T R_1(\theta_U))^N  T   .
\end{eqnarray*}
which satisfies the chiral symmetries introduced above.

The deterministic in- and outcoupling of the light, as already introduced in \cite{nitsche2018probing, nitsche2020local}, minimizes the roundtrip losses with respect to the probabilistic version used earlier \cite{schreiber_photons_2010} and thus enables the recording of longer dynamics.
The detection units consists of two superconducting nanowire single photon detectors (SNSPDs) with dead times well below the pulse separation, which in combination with PBS$_3$ and a QWP at $45^\circ$ enable the measurement of each pulse intensity in the circular $R/L$ basis.

{\it Results:---}We obtain the spin-resolved probability distribution  
by recording measurements for 500 different realisations 
of polarization rotations 
simulating the static binary disorder discussed above. 
To define a reference frame without Anderson localization to compare to, we additionally simulate \textit{dynamic} disorder for which localization is absent due to the destruction of phase coherence. The latter is realized via application of the polarization rotations randomly distributed in time, at otherwise identical system parameters.

The comparison between the two recordings is presented in  Fig.~\ref{Fig:wavefunction}, 
at time step $t=14$, 
traced over polarisation. At these values, the Anderson localized and the diffusively spreading wave package do not yet differ by much in absolute values. Instead, the difference shows in the shape of the distribution. That is, the convex profile of an exponential distribution in the Anderson localized case, compared to the concave form of a Gaussian diffusive profile, as verified in the right panel.
(The boundary peaks in the diffusive case represent a small fraction of quasi-ballistically propagating contributions which disappear in the numerical simulation of longer runs, see Supplemental Material.)

For each random realization of binary angles we extract the spatially integrated spin-polarization,  Eq.~\eqref{eq:integrated_spin_polarization}, which we then average over different angle realizations. 
The upper left panel in Figure~\ref{Fig:spinPol} 
shows  
experimental data,
and corresponding  
numerical simulations,
for $\langle \Delta P(t) \rangle$ between $t=5$ and 14 time steps, 
averaged over 500 realizations of {\em static} binary disorder. 
The upper-right panel exhibits the resulting power spectrum, $S(\omega) = |\bar P(\omega)|^2 / \sum_\omega |\bar P(\omega)|^2$, where $\bar P(\omega) = \sum_{t=5}^{14} e^{i\omega t} \langle \Delta P(t) \rangle $. 
For comparison, we 
show
in the lower left panel 
 the corresponding power spectrum for 
{\em dynamic} binary disorder. 
The latter has a structureless random  
pattern with contributions of the same order 
 from all frequencies, as expected for a noisy spin polarization.
In contrast, the spin polarization 
for static disorder 
 shows the time-staggering predicted 
 for the topological quantum critical state,  
 which is also witnessed by the pronounced peak at $\omega = \pi$ in the power spectrum. 
The relatively small number of time-steps probed in the experiment implies sizeable fluctuations of random parameters, which need to be taken into account in the interpretation of  data: for a given run, the average value of $t_\mathrm{max}$ randomly drawn angular  parameters scales as
 $\sim {\cal O}(1/\sqrt{t})$, and a value of comparable magnitude for the staggered amplitude. 
 In this way, the critical states both, at  $(0,\pi)$ and $(-\pi/2,\pi/2)$ get effectively gapped out for at least a fraction of the runs, resulting in a suppression of peaks in the power spectrum at $\omega=0$ and $\omega=\pi$, respectively (cf. Fig.~\ref{Fig:spinPol}).
To further illustrate this point, 
we repeated numerical simulations of 
 ensembles of 500 random binary angle configurations  
a large number of times. 
In the right-lower panel of Fig.~\ref{Fig:spinPol} we show
the distributions of peak values $S(\omega=0,\pi)$  in the power spectrum,  
resulting  from 1000 repetitions of the previously described procedure.  
That is, each point here presents the average over an ensemble of 500 angle realizations 
(the ensembles used in experiment are indicated by the squares). 
As anticipated above, the distribution for static disordered is centered around large peak values, either at $\omega=0$, at $\omega=\pi$, or at both.
The distribution for dynamic disorder, on the other hand, is always dominated by small peak values at both frequencies. 
The average power spectrum for the entire ensemble of $1000\times 500$  angle configurations (cf. Supplemental Material) then  shows the expected two-peak structure, with peaks at $\omega=0,\pi$ dominating over an otherwise approximately flat background, confirming thus that upon simulating larger runs, fluctuation effects diminish.

{\it Conclusion:---}We have realized an optical linear network simulator of one-dimensional topological quantum criticality.
The simulator implements the photonic  
quantum walk of a spin-1/2 particle 
with chiral symmetry. Its optical network architecture allows us to fully access and  monitor  
the state's internal degree of freedom. Upon tuning to 
the  critical point separating two topological Anderson insulating phases,  
we observe a time-staggered spin-polarization recently suggested as a 
smoking gun signature of quantum critical dynamics. Externally imposed time dependent noise, or `dephasing disorder', destroys the signal, revealing its origin in quantum coherence. A similar destruction takes place upon breaking the chiral symmetry of the walk, and along with it the transition between two symmetry protected topological phases. Ideally, one would like to monitor scaling phenomena induced by such type of symmetry breaking, however, the currently realizable signal times are still too short for such type of statistics. Irrespective of such limitations, we believe that fully programmable quantum networks promise interesting perspectives for the simulation of topological quantum matter in the presence of engineered randomness.

\begin{acknowledgments}
	The Integrated Quantum Optics group acknowledges support by the ERC project QuPoPCoRN (Grant no. 725366). 
  T.~M.~acknowledges financial support by Brazilian agencies CNPq and FAPERJ. A.~A. and D.~B. acknowledge partial support from the Deutsche Forschungsgemeinschaft (DFG) within the CRC network TR 183 (project grant 277101999) as part of projects A01 and A03. K.W.K.~acknowledges financial support by Basic Science Research Program through the National Research Foundation of Korea (NRF) funded by the Ministry of Education (No.2021R1F1A1055797) and Korea government(MSIT) (No.2020R1A5A1016518). 
\end{acknowledgments}



%

\begin{appendix}

\section{Probability distributions at longer times}

\begin{figure*}[t]
\centering 
\includegraphics[width=2\columnwidth] {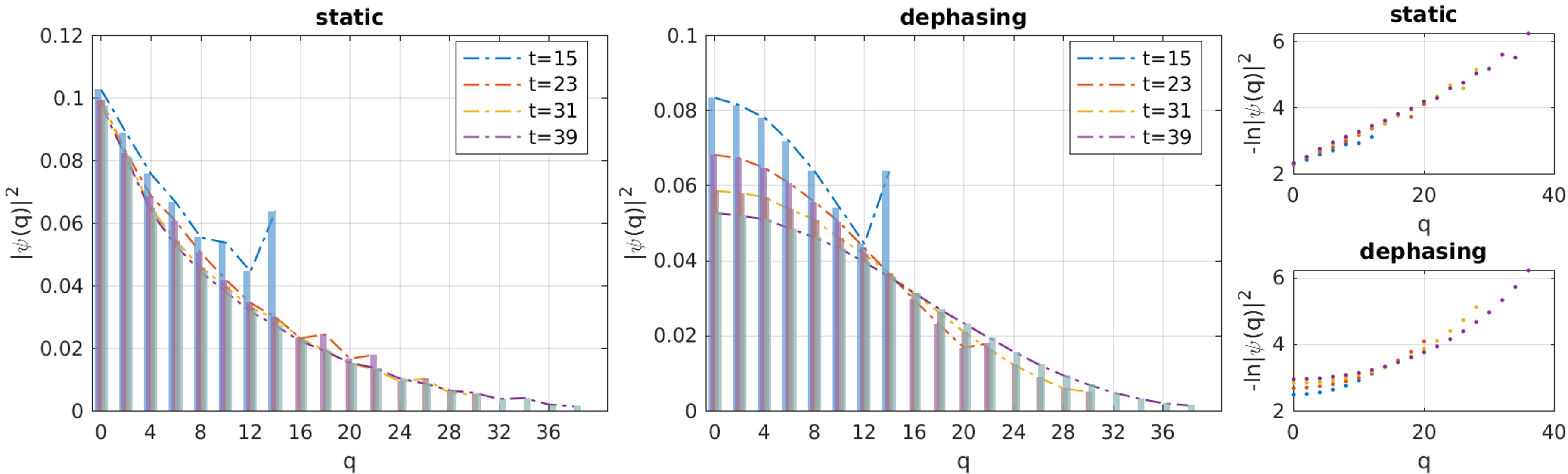}
 \vspace{.4cm}
 \caption{ Probability distributions after $t=15, 23,31,39$ 
 time steps 
 for static (left panel) and dephasing (center panel) disorder. The boundary peak visible at $t=15$ disappears for larger time steps. 
 Right panels show $-\ln(|\psi(q)|^2)$ for static disorder (top) and dephasing disorder (bottom). 
   \label{Fig:append1} 
\vspace{-.2cm} }
\end{figure*}

As discussed in the main text, all states in the quasienergy
spectrum contribute to the probability distribution of the walker. Since most of the states are non-critical, the probability distribution for static binary disorder thus follows  $|\psi(q)|^2\sim e^{-|q|/\lambda}$, as shown in Fig.~\ref{Fig:append1} (left panel) for numerical 
simulations of different time steps $t=15,23,31,39$. 
The $q$-dependence of $-\ln |\psi(q)|^2 $ is plotted in Fig.~\ref{Fig:append1} (right-top panel), clearly showing a  linear dependence on position. 
When the binary disorder also fluctuates in time, quantum
phase coherence is destroyed turning dynamics thus into classical diffusion, $|\psi(q)|^2 \sim e^{-q^2/\sigma_t^2}$, with $\sigma^2_t\sim t$. The numerical simulation,  Fig.~\ref{Fig:append1} (center and right-bottom panel), confirms the diffusive behavior and corresponding scaling of $-\ln |\psi(q)|^2 $. 
Notice that the boundary peaks in the probability distributions are due to a small fraction of 
quasi-ballistically propagating states. 
As shown in 
Fig.~\ref{Fig:append1}, their contribution disappears as longer times $t\gtrsim 20$ are probed.

\section{Statistics of the power spectrum $S(\omega)$}

The relatively small number of time steps measured in the experiment implies that the signature of critical states, viz. the time-staggered spin polarization, is subject to large statistical fluctuations. From our numerical simulations we find that it is necessary to average over many more than 500 disorder realizations to clearly observe the two-peak structure
  in the power spectrum. Figure~3 (bottom-right panel) 
  in the main text shows the distribution of 1000 peak heights at frequencies $\omega=0, \pi$ in the power spectrum
      for static  (purple) and dephasing disorder (blue).
   Each of the 1000 shown points is obtained from averaging the power spectrum over 500 random disorder realizations. 
  The average and standard deviation of the entire power spectrum $S(\omega)$ 
  after averaging over the total ensemble 
  of $1000\times 500$ realizations
  is plotted in Fig.~\ref{Fig:append2} below.

\begin{figure}[t]
\centering
\includegraphics[width=1\columnwidth] {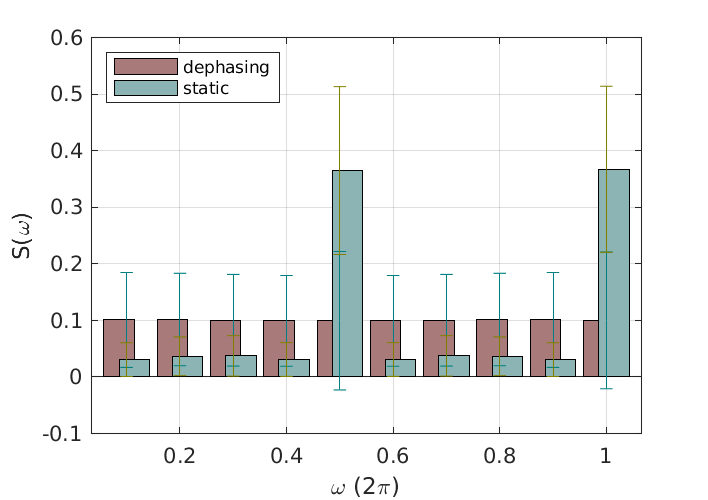}
 \vspace{.4cm}
 \caption{ Average and standard deviation of the power spectrum of the spin polarization for an ensemble of 
 $1000\times 500$ disorder realizations. 
 \label{Fig:append2}
 \vspace{-.2cm} }
\end{figure}

\end{appendix}

\end{document}